\documentclass[pdflatex,sn-mathphys-num]{sn-jnl}

\usepackage{graphicx}%
\usepackage{multirow}%
\usepackage{amsmath,amssymb,amsfonts}%
\usepackage{amsthm}%
\usepackage{mathrsfs}%
\usepackage[title]{appendix}%
\usepackage{xcolor}%
\usepackage{textcomp}%
\usepackage{manyfoot}%
\usepackage{booktabs}%
\usepackage{algorithm}%
\usepackage{algorithmicx}%
\usepackage{algpseudocode}%
\usepackage{listings}%
\usepackage{braket}
\usepackage{bm}

\theoremstyle{thmstyleone}%
\theoremstyle{thmstyletwo}%

\theoremstyle{thmstylethree}%

\begin{document}

\title[Quantum-Classical Neural Networks for Solving Lindblad Master Equations]{Quantum-Classical Neural Networks for Solving Lindblad Master Equations}


\author*[1,4]{\fnm{Rafael} \sur{Florencio}}\email{rfdiaz@ujaen.es}
\equalcont{These authors contributed equally to this work.}

\author[2,4]{\fnm{Julio} \sur{Guerrero}}\email{jguerrer@ujaen.es}
\equalcont{These authors contributed equally to this work.}

\author[3,4]{\fnm{Elvira} \sur{Romera}}\email{eromera@ugr.es}
\equalcont{These authors contributed equally to this work.}

\affil*[1]{\orgdiv{Departamento de Matem\'aticas}, \orgname{Universidad de Ja\'en}, \orgaddress{\street{Edificio de Servicios Generales - Campus Cient\'ifico Tecnol\'ogico de Linares. Avda. de la Universidad (Cintur\'on Sur), s/n}, \city{Linares}, \postcode{23700}, \state{Ja\'en}, \country{Espa\~{n}a}}}

\affil[2]{\orgdiv{Departamento de Matem\'aticas}, \orgname{Universidad de Ja\'en}, \orgaddress{\street{Campus Las Lagunillas}, \city{Ja\'en}, \postcode{23071}, \state{Ja\'en}, \country{Espa\~{n}a}}}

\affil[3]{\orgdiv{Departamento de F\'isica At\'omica, Molecular y Nuclear}, \orgname{Universidad de Granada}, \orgaddress{\street{Avenida de Fuentenueva s/n}, \city{Granada}, \postcode{18071}, \state{Granada}, \country{Espa\~{n}a}}}

\affil[4]{\orgdiv{C\'atedra de Energ\'ia y Tecnolog\'ias Cu\'anticas}, 
	\orgname{Universidad de Granada}, 
	\country{Spain}}

\abstract{\hspace{2em} Variational quantum algorithms have emerged as a promising approach for simulating open quantum systems, yet their expressive power and trainability remain poorly understood. We propose a variational framework for solving Lindblad master equations by minimizing a residual cost function using a compositional ansatz obtained by combining a classical neural network with a parameterized quantum circuit. The density matrix is represented through a purification on an enlarged Hilbert space, ensuring positivity, Hermiticity, and unit trace by construction, while all gradients are evaluated exactly via the parameter-shift rule. Beyond the algorithmic formulation, we derive an upper bound for the covering number of the proposed compositional model and compare it with that of purely quantum neural networks. Our analysis shows that the classical parametrization enlarges the effective hypothesis class while providing finer control over its complexity, thereby establishing a theoretical connection between expressivity, covering numbers, effective dimension, and the barren plateau phenomenon. Numerical experiments validate both the accuracy of the proposed method for representative Lindblad master equations and the theoretical predictions regarding gradient concentration. These results provide a unified variational and theoretical framework for the simulation of dissipative quantum dynamics using compositional quantum machine learning models.}


\maketitle

\section{Introduction}\label{sec1}

\hspace{2em} The interaction of a quantum system with its environment generally gives rise to nonunitary reduced dynamics, which may manifest through dissipation, relaxation, or decoherence. Under the standard assumptions concerning the initial preparation of the system and its environment, this evolution can be described in terms of quantum channels, namely, completely positive trace-preserving linear maps \cite{Breubook,NielBook}. Such channels admit a Kraus-operator representation and, equivalently, a unitary dilation on an enlarged Hilbert space \cite{stinespring1955,kraus1971}.

The form of the reduced dynamics is governed by the timescale over which the environment retains information about the system. When environmental correlations decay on a timescale much shorter than that characterizing the system dynamics, memory effects may be neglected and a Markovian approximation becomes appropriate. If these correlations persist, the evolution also depends on the system's history and may be formulated in terms of integro-differential equations with memory kernels, derived by means of projection-operator techniques \cite{nakajima1958,zwanzig1960}.

In the time-homogeneous Markovian regime, the evolution can be described by a completely positive trace-preserving dynamical semigroup whose generator takes the Gorini--Kossakowski--Sudarshan--Lindblad form \cite{Gori1976,Lind1976}. The corresponding equation, commonly referred to as the Lindblad master equation or the GKSL equation, combines a coherent contribution generated by the Hamiltonian with a dissipator describing processes induced by the system--environment interaction. The superoperator comprising both contributions is termed the Lindblad generator, or Lindbladian, whereas the GKSL equation provides the standard framework for describing Markovian open quantum systems \cite{Lind1975,Breubook,Galk2024}.

These dynamics are central to quantum metrology, energy transport in molecular systems, and dissipative state preparation for applications in quantum computation and quantum information \cite{huelga1997,ishizaki2009,verstraete2009}.

Despite their broad relevance, the numerical treatment of the GKSL master equation rapidly becomes computationally demanding as the system size increases, owing to the exponential growth of both state and operator spaces \cite{NielBook}. This difficulty has motivated the development of compact representations and variational methods that avoid a complete parametrization of the density matrix. Within the quantum-circuit model, an additional structural mismatch arises between the unitary transformations implemented by the circuit and the generally nonunitary evolution of the reduced system. A dilation on an enlarged Hilbert space provides a unitary realization of quantum channels through the introduction of ancillary degrees of freedom followed by a partial trace \cite{stinespring1955,kraus1971,Rast2012}.

The simulation of such dynamics has prompted strategies based on product formulas for local Liouvillians \cite{kliesch2011}, linear-combination techniques adapted to quantum channels and completely positive maps \cite{clevewang2017}, unitary decompositions of nonunitary operators \cite{schlimgen2021}, and variational formulations for mixed states and open-system processes \cite{yuan2019,endo2020}. The choice of ansatz, measurement overhead, numerical stability, and allocation of classical and quantum resources remain central considerations in these methods \cite{delgadogranados2025}.

Recent advances in quantum machine learning have opened new possibilities for representing and simulating quantum systems through parameterized models. Following the success of neural-network quantum states for many-body problems \cite{carleo2017}, parameterized quantum circuits have emerged as expressive function approximators compatible with noisy intermediate-scale quantum hardware. Quantum--classical algorithms, in which a parameterized quantum circuit is optimized through classical routines, have become one of the dominant paradigms in variational quantum computing due to their flexibility and hardware efficiency \cite{Kand2017,Mita2018,TutorialQML}.

These developments have naturally been extended to open quantum systems. Classical neural-network (CNN) representations have been proposed for approximating dissipative many-body dynamics \cite{hartmann2019} and nonequilibrium steady states \cite{vicentini2019}, while deep quantum neural network (QNN)s have recently been employed to solve Lindblad master equations variationally \cite{liu2022}. More generally, differentiable quantum circuits have demonstrated their ability to approximate the solutions of differential equations by minimizing suitably designed residual functions rather than performing direct time integration \cite{kyri2021}. Such residual-based formulations transform the solution of the governing equation into an optimization problem and have become increasingly attractive within variational quantum algorithms.

The effectiveness of these methods, however, depends critically on the expressive power and trainability of the underlying variational ansatz. On the one hand, increasing the complexity of a quantum circuit generally enlarges the family of states that can be represented. On the other hand, highly expressive parameterized quantum circuits may suffer from barren plateaus, where the variance of the gradients decreases exponentially with the number of qubits \cite{McClean2018}. Subsequent studies have shown that this phenomenon strongly depends on the choice of cost function \cite{Cerezo2021} and may also be induced by highly entangled variational states \cite{Marrero2021}. Consequently, understanding the relationship between expressivity and trainability has become one of the central challenges in quantum machine learning.

Several theoretical frameworks have recently been developed to characterize the expressive capabilities of quantum neural networks. For example, dynamical Lie algebras have recently been employed to analyse the expressive power and generalization properties of variational quantum models \cite{ohno2025}. From the perspective of statistical learning theory, concepts such as effective dimension and covering numbers provide quantitative measures of model complexity and approximation capability \cite{MehrBook,ShaiBook}. In particular, covering numbers quantify the minimum number of metric balls required to cover a hypothesis class and play a central role in approximation theory and learning theory \cite{kolmogorov1959,dudley1967,Well2005}. Besides their importance in generalization analysis, covering numbers also provide insight into concentration phenomena in high-dimensional spaces \cite{VersBook}, thereby establishing a theoretical link with the emergence of barren plateaus.

Motivated by these considerations, in this work we propose a variational framework for solving Lindblad master equations by minimizing a residual cost function constructed from physically relevant observables. The density matrix is represented through a purification ansatz on an enlarged Hilbert space, thereby guaranteeing positivity, Hermiticity, and unit trace by construction. Rather than optimizing the rotation angles of the quantum circuit directly, we define them through a classical neural network, so that the resulting variational ansatz is obtained by composing a classical neural network with a parameterized quantum circuit. In this way, the classical model generates a smooth parametrization of the quantum circuit, while the latter maps these parameters into a purified quantum state. This compositional construction combines the expressive capabilities of quantum neural networks with the flexibility of classical function approximation. Furthermore, all gradients of the proposed residual cost function can be evaluated exactly by means of the parameter-shift rule \cite{Schu2019}, making the method compatible with current quantum hardware.

Beyond the algorithmic formulation, we investigate how this compositional architecture influences both the expressive power and the trainability of the resulting variational model. To this end, we derive an upper bound for its covering number and compare it with the corresponding bound for a purely quantum neural network. Our analysis shows that composing a parameterized quantum circuit with a classical neural network enlarges the effective hypothesis class while providing finer control over its complexity through the classical parametrization. Building upon these results, we establish a connection between covering numbers, the effective dimension of the variational manifold, and the barren plateau phenomenon, thereby providing a theoretical explanation of the trainability properties of the proposed model. Finally, numerical experiments illustrate both the capability of the proposed approach to solve representative Lindblad master equations and the theoretical predictions concerning gradient concentration and optimization behaviour.

\section{Minimizing Lindbland's Residual Using an Ansatz made of QNN composed by CNN}\label{sec2}

\hspace{2em}Consider an open quantum system $S$ with a finite-dimensional Hilbert space 
$\mathcal{H}_S$ of dimension $2^{K_S}$. The general master equation governing the time 
evolution of the density matrix $\rho$, which preserves both trace and positivity, can 
be written in the Lindblad form (setting $\hbar = 1$) \cite{Lind1975,Gori1976}.

\begin{align}
	\label{eq1}
	&\dot{\rho} = \mathcal{L}(\rho)
\end{align}

where $\dot{\rho}=\frac{d\rho}{dt}$ and where:
\begin{align}
	\label{eq2}
&\mathcal{L}(\rho)= -i[H, \rho] + \sum_{k=1}^{2K_S} \gamma_k \left( L_k \rho L_k^\dagger - \frac{1}{2} \{ L_k^\dagger L_k, \rho \} \right)
\end{align}

Note that while the Hilbert-space description of a pure state involves $2^{K_S}$ degrees of freedom, solving the Lindblad master equation requires evolving the full 
density matrix, whose number of independent entries scales as $2^{2K_S}$. To solve the master equation without explicitly computing this large number of unknowns, we 
propose minimizing its residual by means of a quantum neural network (QNN) composed with a classical neural network (CNN), using a residual defined from the discrepancy between predicted and physical observables.

Consider an enlarged Hilbert space $\mathcal{H}_S \otimes \mathcal{H}_A$, where $S$ denotes the physical system and $A$ an auxiliary (anchor) system. We prepare the variational state generated by the QNN as follows:
\begin{align}
	\label{eq3}
	&\ket{\Psi(\bm{\theta})} = U(\bm{\theta}) \left(\ket{0}_S\otimes\ket{0}_A\right)
\end{align}

where $\ket{0}_S=\ket{0}^{\otimes K_S}$ and $\ket{0}_A=\ket{0}^{\otimes K_A}$, with $\ket{0}=(1,0)^T$ and $(\cdot)^T$ denoting the matrix transpose. The unitary operator $U$ is composed of rotation gates $U_j(\theta_j)$ with rotation angle $\theta_j$ and entangling gates $U_{\text{ent}}$ (e.g., CNOT gates). Consider $L$ layers, a fixed number $n_\mathrm{rgq}$ of rotation gates per qubit in each layer (or per wire in the quantum-circuit context), $k$ wires, and the parameter vector $\bm{\theta}=(\theta_1,\ldots,\theta_{N_\mathrm{rg}})$ with $N_\mathrm{rg}=n_\mathrm{rgq}\, k\, L$ the total number of rotation gates. One possible construction of the operator $U$ is given in \cite{Kand2017,kyri2021}:
\begin{align}
	\label{eq4}
	&U(\bm{\theta}) = \prod_{l=1}^{L} \left[ U_{\text{ent}} \cdot \left( \bigotimes_{j=1}^{n_\mathrm{rgq}\cdot k} U_j(\theta_{j+n_\mathrm{rgq}\cdot k\cdot(l-1)}) \right) \right]
\end{align}

\begin{figure}[H]
	\centering
	\includegraphics [width=14 cm]{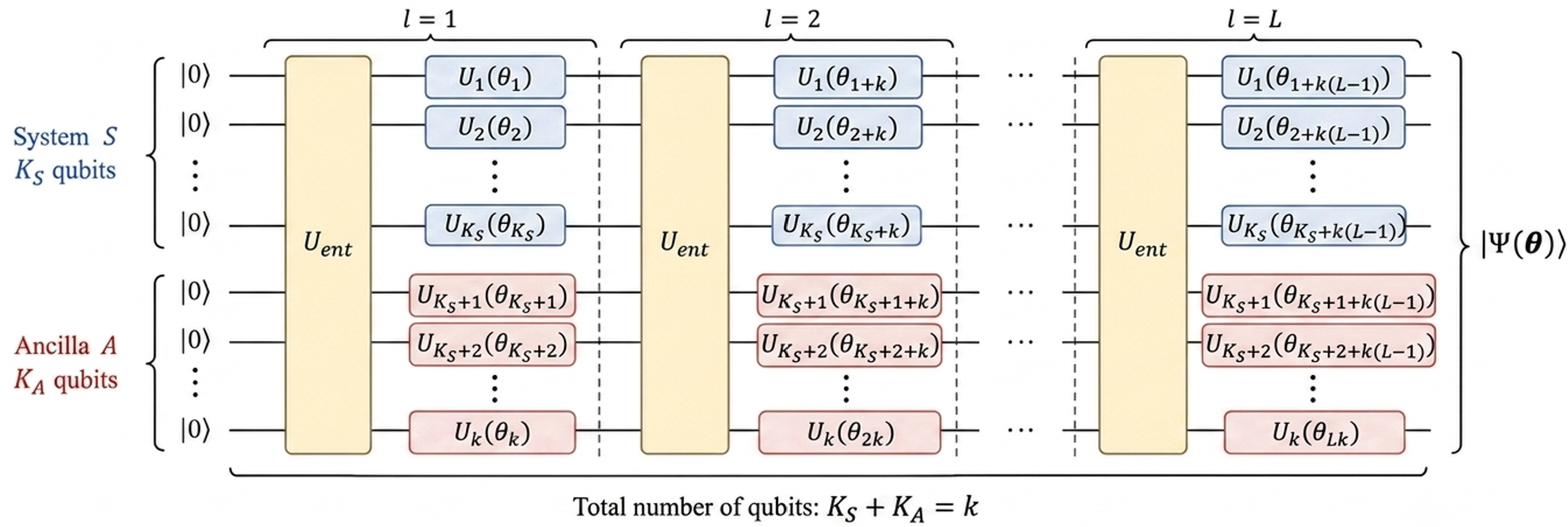}
	\caption{\label{fig1} Schematic representation of the quantum neural network (the case $n_{rgq}=1$ is shown for simplicity)}
\end{figure}

A schematic representation of the QNN is shown in Fig.~\ref{fig1}.

The density matrix $\rho$ is obtained by taking the partial trace over the ancilla Hilbert space $\mathcal{H}_A$ of the projector onto the state:
\begin{align}
\label{eq5}
\rho(\bm{\theta}) = \text{Tr}_A \left( \ket{\Psi(\bm{\theta})}\bra{\Psi(\bm{\theta})} \right)
\end{align}

This approach naturally ensures hermiticity, positivity ($\rho \ge 0$), and the unit-trace condition of the density operator, without the need to impose additional 
constraints during optimization.

The key idea of our proposal is to replace static variational parameters with a dynamical trajectory in Hilbert space, parameterized by a CNN. The goal is to model the quantum-circuit parameter vector $\bm{\theta}$ using a classical neural network that takes time $t$ as its single input, equipped with trainable weights $w_{j\alpha}$ and bias terms $b_{j\alpha}$ (which we collectively store in the vector $\mathbf{w}$), and with $N_{\mathrm{rg}}$ output components corresponding to the entries of $\bm{\theta}$:
\begin{align}
	\label{eq6}
	&\bm{\theta}(t; \mathbf{w}) = \bm{\theta}_0+t\bm{N}_c(t; \mathbf{w})
\end{align}

where $\bm{\theta}_0 = (\theta_{0,1}, \ldots, \theta_{0,N_{\mathrm{rg}}})$ denotes the initial rotation angles, chosen so as to fix the initial density matrix of the system.

In this work, we use a single-layer neural network whose $j$-th output is defined as follows:
\begin{align}
	\label{eq7}
	&\left(\bm{N}_c(t; \mathbf{w})\right)_j=\sum_{\alpha=1}^{n_{w,j}}\tanh\left(w_{j\alpha}t+b_{j\alpha}\right)
\end{align}

\begin{figure}[H]
	\centering
	\includegraphics [width=10 cm]{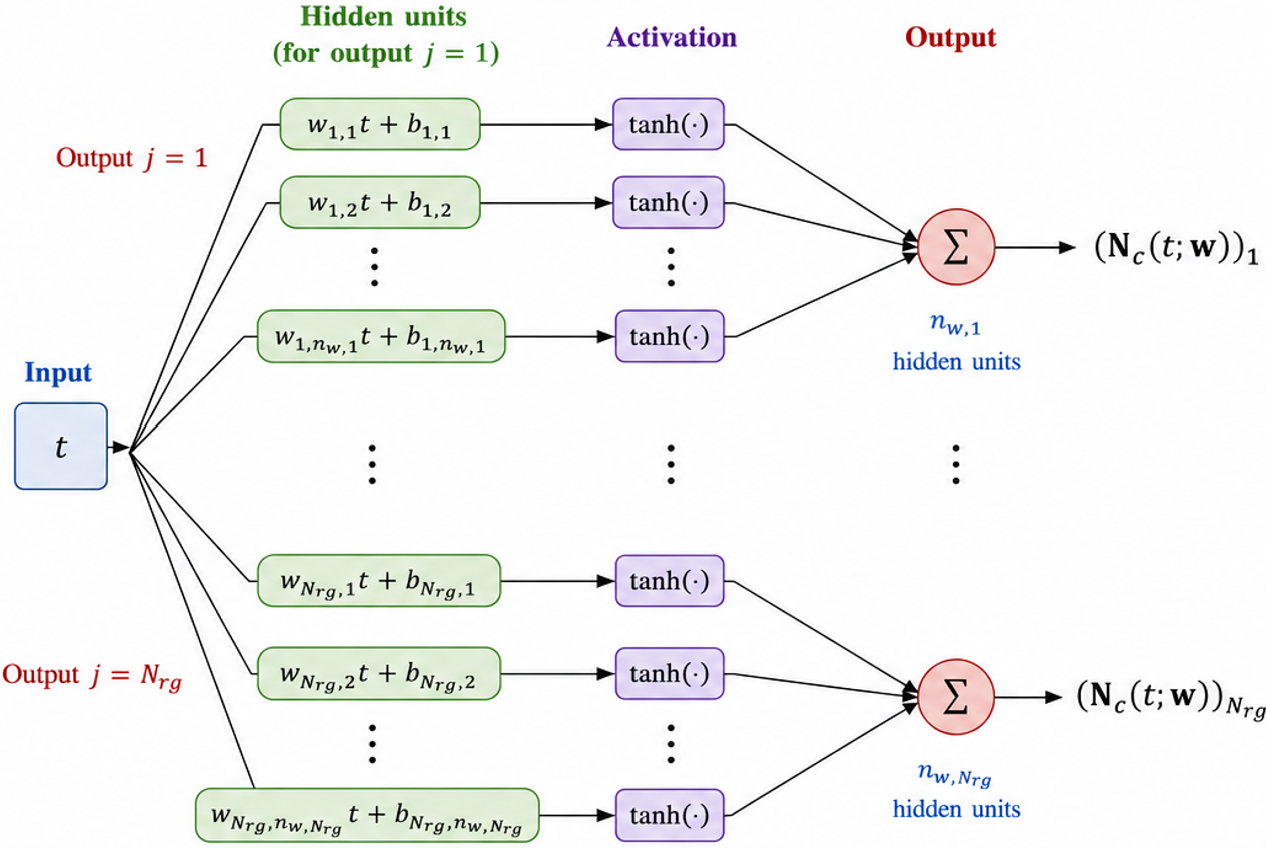}
	\caption{\label{fig2} Classical neural network schematic.}
\end{figure}

A schematic of the classical neural network is shown in Fig.~\ref{fig2}. The total number of parameters of the CNN is denoted by $n_{w}=2\sum_{j=1}^{N_{\mathrm{rg}}} n_{w,j}$. In Section~3, we study the expressivity of the proposed QNN-CNN composition in terms of $n_{w}$ and $N_{\mathrm{rg}}$.

Consider a residual cost function defined with respect to a basis of Hermitian witness operators $\{O_a\}_{a=1}^M$ (from now on referred to as the Lindblad residual cost function):
\begin{align}
	\label{eq8}
	&\mathcal{C}(t; \mathbf{w}) = \sum_{a=1}^M  \left[\text{Tr}(O_a \dot{\rho}) - \text{Tr}(O_a \mathcal{L}(\rho)) \right]^2
\end{align}

The basis of Hermitian witness operators $\{O_a\}_{a=1}^M$ must be strategically selected to capture the physical evolution of the system (i.e., the Hamiltonian $H$ 
and the Lindblad operators $L_k$), while enabling the detection of faulty components in the QNN. For example, if both $H$ and the $L_k$ operators involve only the Pauli  operator $\sigma_z$, and the QNN introduces an unwanted $\sigma_x$ component, such a component will only be detected if $\sigma_x$ is included in $\{O_a\}_{a=1}^M$. In this work, we consider $\{O_a\}_{a=1}^M$ as the set consisting of the Pauli operators together with the identity operator.

To minimize the Lindblad residual cost function defined in \eqref{eq8}, we apply 
gradient descent:
\begin{align}
	\label{eq9}
	&w_{\beta}\leftarrow w_{\beta}-\eta\frac{\partial \mathcal{C}}{\partial w_\beta}
\end{align}

where $\eta$ is the learning rate, and a single index $\beta$ is used to generically refer to any element of the global set of weights and biases in $\mathbf{w}$, in order to simplify the development of gradients in subsequent sections and avoid index overload. According to the definition of the residual cost function, the gradient of this residual with respect to the weights and biases of the CNN is given by:
\begin{align}
	\label{eq10}
	&\frac{\partial \mathcal{C}}{\partial w_\beta} = 2 \sum_a \left(\text{Tr}(O_a \dot{\rho}) - \text{Tr}(O_a \mathcal{L}(\rho)) \right) \left( \frac{\partial \text{Tr}(O_a \dot{\rho})}{\partial w_\beta} - \frac{\partial \text{Tr}(O_a \mathcal{L}(\rho))}{\partial w_\beta} \right)
\end{align}

Note that the computation of $\frac{\partial \mathcal{C}}{\partial w_\beta}$ involves the gradients $\frac{\partial \mathrm{Tr}(O_a \dot{\rho})}{\partial w_\beta}$ and $\frac{\partial \mathrm{Tr}(O_a \mathcal{L}(\rho))}{\partial w_\beta}$. These gradients can be obtained by means of the parameter-shift rule (PSR) \cite{Mita2018,Schu2019}. In the next subsection, we show the explicit computation of these gradients using the chain rule and the PSR.

\subsection{Gradients of the Lindblad residual cost function via the parameter-shift rule}
\label{subsec21}

\hspace{2em}In this subsection we show the explicit computation of the terms $\frac{\partial \text{Tr}(O_a \dot{\rho})}{\partial w_\beta}$ and $\frac{\partial \text{Tr}(O_a \mathcal{L}(\rho))}{\partial w_\beta}$ of the gradients of the Lindblad residual cost function given in \eqref{eq10} via PSR.
\\

Consider the operator $\mathcal{L}^\dagger(O_a)$ in the Heisenberg picture, defined as \cite{Breubook, Galk2024}: 
\begin{align}
	\label{eq11}
	&\text{Tr}\big(O_a \mathcal{L}(\rho)\big) = \text{Tr}\big(\mathcal{L}^\dagger(O_a) \rho\big)
\end{align}

where $\mathrm{Tr}\!\big(\mathcal{L}^\dagger(O_a)\rho\big)=\langle \mathcal{L}^\dagger(O_a) \rangle$ 
denotes the expectation value of the operator $\mathcal{L}^\dagger(O_a)$, and where
\begin{align}
	\label{eq12}
	\mathcal{L}^\dagger(O_a) = i[\hat{H}, O_a] + \sum_{k=1}^{2K_S} \gamma_k \left( \hat{L}_k^\dagger O_a \hat{L}_k - \frac{1}{2}\{\hat{L}_k^\dagger \hat{L}_k, O_a\} \right)
\end{align}

The resultant operator $\mathcal{L}^\dagger(O_a)$ is Hermitian whenever $O_a$ is Hermitian. Consider the set consisting of the Pauli operators $\{P_r\}$ together with the identity operator $I$. Since this set forms a basis for the space of Hermitian operators \cite{NielBook}, any Hermitian operator $O_a$ implies that $\mathcal{L}^\dagger(O_a)$ can be expanded in this basis as
\begin{align}
	\label{eq13}
	&\mathcal{L}^\dagger(O_a)=c_I I+\sum_{r}c_r P_r
\end{align}

Using \eqref{eq11} and \eqref{eq13}, the computation of $\frac{\partial \mathrm{Tr}(O_a \mathcal{L}(\rho))}{\partial w_\beta}$ yields:
\begin{align}
	\label{eq14}
	&\frac{\partial \text{Tr}(O_a \mathcal{L}(\rho))}{\partial w_\beta}=\sum_r\sum_j c_r\frac{\partial \langle P_r \rangle}{\partial \theta_j}\frac{\partial \theta_j}{\partial w_\beta}
\end{align}

where $\langle P_r \rangle = \mathrm{Tr}(P_r \rho)$ has been used. Because each Pauli operator $P_r$ has eigenvalues $\pm 1$, the requirements for the PSR are fulfilled \cite{Mita2018,Schu2019}. Hence, the derivatives $\frac{\partial \langle P_r \rangle}{\partial \theta_j}$ appearing in \eqref{eq14} 
can be computed exactly using: 
\begin{align}
	\label{eq15}
	&\frac{\partial \langle P_r \rangle}{\partial \theta_j}=\frac{1}{2}\left(\langle P_r \rangle_{\theta_j+\frac{\pi}{2}}-\langle P_r \rangle_{\theta_j-\frac{\pi}{2}}\right)
\end{align}

where $\langle P_r \rangle_{\theta_j+\frac{\pi}{2}}$ and $\langle P_r \rangle_{\theta_j-\frac{\pi}{2}}$ denote the expectation values of the operator $P_r$ evaluated at ansatz parameters in which $\theta_j$ is shifted by $\pm \frac{\pi}{2}$.

Because the trace and the time derivative commute, 
$\mathrm{Tr}(O_a \dot{\rho})$ can be written as the time derivative of the expectation value of the operator $O_a$. Using the chain rule, we obtain:
\begin{align}
	\label{eq16}
	&\text{Tr}(O_a\dot{\rho})=\sum_{j} \frac{\partial \langle O_a \rangle}{\partial \theta_j} \frac{d\theta_j}{dt}
\end{align}

We next compute the partial derivative of \eqref{eq16} with respect to the classical weight $w_\beta$. Using the chain rule, we obtain:
\begin{align}
	\label{eq17}
	&\frac{\partial \text{Tr}\big(O_a \dot{\rho}\big)}{\partial w_\beta} = \sum_{i=1}^{N}\sum_{j=1}^{N} \frac{\partial^2 \langle O_a \rangle}{\partial \theta_j \partial \theta_i} \frac{d\theta_i}{dt} \frac{\partial \theta_j}{\partial w_\beta} + \sum_{i=1}^{N} \frac{\partial \langle O_a \rangle}{\partial \theta_i} \frac{\partial^2 \theta_i}{\partial w_\beta \partial t}
\end{align}

If $O_a$ is chosen to be a Pauli operator, then both 
$\frac{\partial^2 \langle O_a \rangle}{\partial \theta_j \partial \theta_i}$ and 
$\frac{\partial \langle O_a \rangle}{\partial \theta_i}$ can be obtained by applying 
the PSR once more. This yields:
\begin{align}
	\label{eq18}
	&\frac{\partial \langle O_a \rangle}{\partial \theta_i} = \frac{1}{2} \left( \langle O_a \rangle_{\theta_i + \frac{\pi}{2}} - \langle O_a \rangle_{\theta_i - \frac{\pi}{2}} \right)
\end{align}
\begin{align}
	\label{eq19}
	&\frac{\partial^2 \langle O_a \rangle_{\theta_i}}{\partial \theta_j \partial \theta_i}= \frac{1}{4} \Big( \langle O_a \rangle_{\theta_i + \frac{\pi}{2}, \theta_j + \frac{\pi}{2}} - \langle O_a \rangle_{\theta_i - \frac{\pi}{2}, \theta_j + \frac{\pi}{2}} \\
	\nonumber
	&-\langle O_a \rangle_{\theta_i + \frac{\pi}{2}, \theta_j - \frac{\pi}{2}} + \langle O_a \rangle_{\theta_i - \frac{\pi}{2}, \theta_j - \frac{\pi}{2}} \Big)
\end{align}

The parameter-shift rule provides a substantial acceleration in the computation of these gradients. By substituting numerical differentiation with a small number of circuit evaluations at shifted parameters, the method produces exact derivatives and reduces both runtime and computational overhead.

\section{Expressivity and plateau barren problem}\label{sec3}

\hspace{2em} Expressivity refers to a model's ability to represent a wide range of functions, determining the smallest achievable training error. We characterize the expressiveness of the composition of a QNN and a CNN using the covering number, an 
advanced tool from statistical learning theory. In mathematics, the covering number $\mathcal{N}(\varepsilon,\mathcal{H}_S)$ is defined as the minimum number of balls of radius $\varepsilon>0$ required to cover a space $\mathcal{H}_S$, allowing overlaps between balls \cite{ShaiBook,MehrBook}. The covering number quantifies the size of a set and applies to general metric spaces. In our case, we aim to characterize the size of the metric space induced by our model, based on the composition of a QNN 
and a CNN. To this end, we recall the following properties of covering numbers:
\\

\textbf{Property 1.\footnote{See Lemma 27.3 in \cite{ShaiBook}.}} 
\emph{Let $f$ be a Lipschitz-continuous function with Lipschitz constant $L_f$ defined on a set $A$. Then, for any $\varepsilon > 0$, the covering number of the  image $f(A)$ is upper bounded as follows:}
\begin{align}
	\label{eq20}
	&\mathcal{N}(\varepsilon,f\circ A)\leq \mathcal{N}(\frac{\varepsilon}{L_f},A)
\end{align}

This property will be instrumental in deriving an upper bound on the covering number of the composition of Lipschitz-continuous functions.
\\

\textbf{Property 2.\footnote{See Lemma 2.2(a) in \cite{Balo1988}.}} 
\emph{Let $A = A_1 \times A_2 \times \cdots \times A_N$ be a subset of a Euclidean space. Then, for any $\varepsilon > 0$, the covering number of $A$ is upper bounded as follows:}
\begin{align}
	\label{eq21}
	&\mathcal{N}(\varepsilon, A)\leq \prod_{i=1}^N\mathcal{N}(\frac{\varepsilon}{\sqrt{N}},A_i)
\end{align}

This property is employed, for example, in \cite{Kara2025,Well2005}. In our setting, we make use of a specialized formulation applicable to subspaces endowed with the Euclidean norm.
\\

\textbf{Property 3.\footnote{See Example 27.1 in \cite{ShaiBook}.}} 
\emph{Let $A \subset \mathbb{R}^N$ be a subset endowed with the Euclidean norm, and let $C = \max_{\mathbf{a} \in A} \|\mathbf{a}\|$. Then, for any $\varepsilon > 0$, the covering number of $A$ admits the following upper bound:}
\begin{align}
	\label{eq22}
	&\mathcal{N}(\varepsilon,A)\leq \left(\frac{2C\sqrt{N}}{\varepsilon}\right)^{N}
\end{align}

Using Properties 1, 2, and 3, we can derive the following theorem:
\\

\textbf{Theorem 1.} 
\emph{
	Let $f = \mathrm{Tr}\!\left(\mathrm{Tr}_A(\rho_T)\, O \right)$, where $\rho_T = \ket{\Psi(\bm{\theta})}\bra{\Psi(\bm{\theta})}$ and $O$ is a generic observable. The state $\ket{\Psi(\bm{\theta})}$ is defined in \eqref{eq3} and \eqref{eq4}, and the parametrization $\bm{\theta}(t;\mathbf{w})$ is defined in \eqref{eq6} and	\eqref{eq7}. 
	Let $k = K_S + K_A$ be the total number of qubits used to define $\ket{\Psi(\bm{\theta})}$, $N_\mathrm{rg}$ the total number of rotating gates, and $n_w$ the total number of CNN parameters. Let $w_{\max} = \max_{\mathbf{w}} \|\mathbf{w}\|$ denote the maximum norm of the CNN parameter vector, and let $L_w$ be the Lipschitz constant of $\bm{\theta}(\mathbf{w})$. Then, for any $\varepsilon > 0$, the covering number of the mapping $\mathbf{w} \mapsto f(\rho_T(\bm{\theta}(\mathbf{w})))$ is upper bounded as follows:} 
\begin{align}
	\label{eq23}
	&\mathcal{N}(\varepsilon, f\circ\rho_T\left(\bm{\theta}\left( \mathbf{w}\right)\right)) \leq \left( \frac{2 w_{\max} \sqrt{n_w} 2^{K_A/2}\|O\|_F 2^k N_\mathrm{rg} L_w}{\varepsilon} \right)^{n_w \cdot N_\mathrm{rg} \cdot 2^{2k}}
\end{align}

\begin{proof}[Proof of Theorem 1]
	Let us equip the space of matrix representations of the operators with the Frobenius norm $\|\cdot\|_F$. Using the upper bound for the Frobenius norm of the partial trace, $\|\mathrm{Tr}_A(X)\|_F \leq \sqrt{d_A}\,\|X\|_F$ \cite{Rast2012}, together with the Cauchy-Schwarz inequality, it is straightforward to verify by 
	direct calculation that the Lipschitz constant of $f$ is given explicitly by $L_f = 2^{K_A/2}\|O\|_F$.
	 
	We first apply \textbf{Property 1} to the function $f$ endowed with the Frobenius norm, obtaining:	
	\begin{align}
		\label{eq24}
		&\mathcal{N}(\varepsilon, f\circ\rho_T\left(\bm{\theta}\left( \mathbf{w}\right)\right)) \leq \mathcal{N}\left(\frac{\varepsilon}{2^{K_A/2}\|O\|_F}, \rho_T\right)
	\end{align}
	The total density matrix $\rho_T$ lies in a product space spanned by its $2^{2k}$ components $\rho_{\alpha\beta}$. By applying \textbf{Property 2} to this Euclidean space, we obtain that its covering number is upper bounded by the product of the covering numbers of its components:
	\begin{align}
		\label{eq25}
		\nonumber
		\mathcal{N}\Big(\frac{\varepsilon}{2^{K_A/2}\|O\|_F}&, \rho_T\Big) \leq  \prod_{\alpha=1}^{2^k} \prod_{\beta=1}^{2^k} \mathcal{N}\left(\frac{\varepsilon}{2^{K_A/2}\|O\|_F 2^k}, \rho_{\alpha\beta}\right) = \\
		&\left[ \mathcal{N}\left(\frac{\varepsilon}{2^{K_A/2}\|O\|_F 2^k}, \rho_{\max}\right) \right]^{2^{2k}}
	\end{align}
	where $\rho_{\max} = \max\{\rho_{\alpha\beta}\}_{\alpha,\beta=1}^{2^k}$.Each component of the density matrix is given by the bilinear relation $\rho_{\alpha\beta} = \psi_\alpha^* \psi_\beta$, where $\psi_{\alpha} = \langle e_{\alpha} | \Psi \rangle$ for some element of the orthonormal basis $\{\ket{e_{\alpha}}\}_{\alpha=1}^{2^k}$. From this mapping, it is straightforward to verify by direct calculation that the Lipschitz constant of the mapping is $L_\rho = 2$. By applying \textbf{Property 1}, we bound the covering number of each individual component by the covering number of the vector of expansion coefficients $\psi$:
	\begin{align}
		\label{eq26}
		&\mathcal{N}\left(\frac{\varepsilon}{2^{K_A/2}\|O\|_F 2^k}, \rho_{\alpha\beta}\right) \leq \mathcal{N}\left(\frac{\varepsilon}{2 \cdot 2^{K_A/2}\|O\|_F 2^k}, \psi\right)
	\end{align}
	The vector of expansion coefficients $\psi$ is prepared as a function of the angle vector $\bm{\theta}\in\mathbb{R}^{N_\mathrm{rg}}$ by means of the operator $U(\bm{\theta})$ given in \eqref{eq4}. The Lipschitz constant for this mapping with respect to the angles under the Euclidean norm is $L_\psi = \frac{\sqrt{N_\mathrm{rg}}}{2}$. To rigorously derive this Lipschitz constant $L_\psi$, we consider two parameter vectors $\bm\theta, \bm\eta \in \mathbb{R}^{N_\mathrm{rg}}$. Let 
	$\psi_{\alpha}(\bm{\theta})=\bra{e_{\alpha}}U(\bm{\theta})\ket{0}$ and $\psi_{\alpha}(\bm{\eta})=\bra{e_{\alpha}}U(\bm{\eta})\ket{0}$, where $\langle e_{\alpha}|e_{\beta}\rangle = \delta_{\alpha,\beta}$ and $\langle \Psi|\Psi\rangle = 1$. By the Cauchy-Schwarz inequality and applying a telescopic expansion, we obtain:
	\begin{align}
		\label{eq27}
		\nonumber
		|\psi_\alpha(\boldsymbol\theta) - \psi_\alpha(\boldsymbol\eta)|^2=&|\bra{e_{\alpha}}U(\bm{\theta})-U(\bm{\eta})\ket{0}|^2 \leq \|U(\boldsymbol\theta) - U(\boldsymbol\eta)\|^2\\
		\nonumber
		&\leq \sum_{r=1}^{N_\mathrm{rg}} \|U_r(\theta_r) - U_r(\eta_r)\|^2 \\
		&\leq \sum_{r=1}^{N_\mathrm{rg}} \frac{1}{2}|\theta_r - \eta_r|^2
		\leq \frac{\sqrt{N_\mathrm{rg}}}{2} \|\boldsymbol\theta - \boldsymbol\eta\|^2_2
	\end{align}
	
	where $\|\cdot\|$ and $\|\cdot\|_2$ denote the spectral and Euclidean norms, respectively, and where the form $U_r(\theta_r) = e^{-i\theta_r P_r/2}$ has been taken into account (with $P_r^2 = I$, which implies $\|\partial_{\theta_r} U_r\| = \tfrac{1}{2}$ since $U_r(\theta_r)$ is unitary). This establishes that $L_\psi =\tfrac{\sqrt{N_\mathrm{rg}}}{2}$.

	Applying \textbf{Property 1} once again enables us to transfer the radius of the balls to the angle vector space:
	\begin{align}
		\label{eq28}
		&\mathcal{N}\left(\frac{\varepsilon}{2 \cdot 2^{K_A/2}\|O\|_F 2^k}, \psi\right) \leq \mathcal{N}\left(\frac{\varepsilon}{2^{K_A/2}\|O\|_F 2^k \sqrt{N_\mathrm{rg}}}, \bm{\theta}\right)
	\end{align}
	
	Next, the parameter space of angles $\bm{\theta}$ is a product space of $N_\mathrm{rg}$ individual angles $\theta_i$. Applying \textbf{Property 2} to this Euclidean space, the upper bound on the covering number decomposes into the product of the covering numbers of the individual angles, with the radius scaled by $1/\sqrt{N_\mathrm{rg}}$:
	
	\begin{align}
		\label{eq29}
		\nonumber
		\mathcal{N}\big(\frac{\varepsilon}{2^{K_A/2}\|O\|_F 2^k \sqrt{N_\mathrm{rg}}}, &\bm{\theta}\big) \leq \prod_{i=1}^{N_\mathrm{rg}} \mathcal{N}\left(\frac{\varepsilon}{2^{K_A/2}\|O\|_F 2^k \sqrt{N_\mathrm{rg}} \sqrt{N_\mathrm{rg}}}, \theta_i\right) =\\ 
		&\hspace{-1.2cm}\left[ \mathcal{N}\left(\frac{\varepsilon}{2^{K_A/2}\|O\|_F 2^k N_\mathrm{rg}}, \theta_{\max}\right) \right]^{N_\mathrm{rg}}
	\end{align}
	where $\theta_{\max} = \max\{\theta_i\}_{i=1}^{N_{\mathrm{rg}}}$. Each individual angle $\theta_i$ is produced by a classical neural network as a function of the weight vector $\mathbf{w}$, with Lipschitz constant denoted by $L_w$. Applying  \textbf{Property 1}, we bound the covering number of each angle by the covering number of the weight space:
	\begin{align}
		\label{eq30}
		&\mathcal{N}\left(\frac{\varepsilon}{2^{K_A/2}\|O\|_F 2^k N_\mathrm{rg}}, \theta_{\max}\right) \leq \mathcal{N}\left(\frac{\varepsilon}{2^{K_A/2}\|O\|_F 2^k N_\mathrm{rg} L_w}, \mathbf{w}\right)
	\end{align}
	The weight vector $\mathbf{w}$ belongs to a bounded subspace of $\mathbb{R}^{n_w}$ equipped with the Euclidean norm, where $w_{\max} = \max_{\mathbf{w}} \|\mathbf{w}\|$. Applying \textbf{Property 3} with the fully accumulated internal radius $\varepsilon' = \frac{\varepsilon}{2^{K_A/2}\|O\|_F\, 2^k\, N_{\mathrm{rg}}\, L_w}$, we obtain the explicit upper bound for the weight space:
	\begin{align}
		\label{eq31}
		&\mathcal{N}\left(\frac{\varepsilon}{2^{K_A/2}\|O\|_F 2^k N_\mathrm{rg} L_w}, \mathbf{w}\right) \leq \left( \frac{2 w_{\max} \sqrt{n_w} 2^{K_A/2}\|O\|_F 2^k N_\mathrm{rg} L_w}{\varepsilon} \right)^{n_w}
	\end{align}
	Finally, by combining all inequalities along the chain of accumulated exponents, we arrive at the final upper bound for the total covering number of the composite hybrid model:	
	\begin{align}
		\label{eq32}
		&\mathcal{N}(\varepsilon, f\circ\rho_T\left(\bm{\theta}\left( \mathbf{w}\right)\right)) \leq \left[ \left[ \left( \frac{2 w_{\max} \sqrt{n_w} 2^{K_A/2}\|O\|_F 2^k N_\mathrm{rg} L_w}{\varepsilon} \right)^{n_w} \right]^{N_\mathrm{rg}} \right]^{2^{2k}}
	\end{align}
	This yields the final expression stated in \eqref{eq23}.
\end{proof}

Let us compare the obtained upper bound on the covering number of our model with the covering number of a pure QNN provided in \cite{TutorialQML}, where 
Theorem~4.13 states the following:
\\

\textbf{Theorem 4.13 of \cite{TutorialQML}.} 
\emph{Suppose that the employed QNN contains in total $N_{\mathrm{rg}}$ training gates, each training gate acting on at most $k$ qubits. Then, for all $\varepsilon > 0$, the covering number of the model for measurements with respect to an observable $O$ obeys}:

\begin{align}
	\label{eq33}
	&\mathcal{N}(\varepsilon,f\circ\rho_T\left(\bm{\theta}\right))\leq \left(\frac{7N_\mathrm{rg}\|O\|}{\epsilon}\right)^{2^{2k}N_\mathrm{rg}}
\end{align}
\\

The dependence of the upper bound in \eqref{eq23} on the number of parameters of our model is 
$\mathcal{O}\!\left(\left(\sqrt{n_w}\,2^{K_A/2}\,2^{k}\,N_{\mathrm{rg}}\right)^{\,n_w N_{\mathrm{rg}} 2^{2^{k}}}\right)$, whereas the dependence of the upper bound in \eqref{eq33} is $\mathcal{O}\!\left(N_{\mathrm{rg}}^{\,N_{\mathrm{rg}} 2^{2^{k}}}\right)$. Ideally, we may regard our model as operating on an effective manifold whose dimension is larger by a factor of $n_w$ than the effective manifold associated with a pure QNN. This indicates that the number of degrees of freedom of a pure QNN can be expanded by a factor of $n_w$, as if the classical parameters allowed the quantum parameter space to be smoothly deformed (for instance, from modeling with straight lines to modeling with spline functions).

From another perspective, the number of degrees of freedom of our model can be controlled linearly through the number of classical parameters $n_w$, whereas in a pure QNN the number of degrees of freedom grows exponentially and uncontrollably with the number of qubits. A similar interpretation applies to the 
maximum distance in the effective manifold: in our model, this maximum distance scales by a factor $\sqrt{n_w}\,2^{K_A/2}\,2^{k}$ relative to the effective manifold of a pure QNN. Consequently, the number of functions that can be represented explodes as the number of qubits increases, but grows sublinearly (and with greater controllability) as the number of classical parameters increases.

These observations show that fine-tuning of our model should be carried out primarily by increasing the number of classical parameters $n_w$, while the 
number of qubits serves to expand the landscape exponentially. Thus, the following practical recipe should be followed when training our model:
\begin{itemize}
	\item First, fix a small number of qubits and increase the number of classical parameters $n_w$ of the neural network until the desired behaviour 
	is captured.
	\item If the model still fails to reproduce the target behaviour, gently increase the number of qubits and repeat the first step.
\end{itemize}

From the behaviour of the covering number of pure QNNs, a similar strategy might seem applicable to the number of training gates $N_{\mathrm{rg}}$. However, we must recall that $N_{\mathrm{rg}}$ increases linearly with the number of qubits $k$ according to $N_{\mathrm{rg}} = n_{\mathrm{rgq}} \cdot k \cdot L$, which couples these parameters and limits such an approach.
\\

From \eqref{eq23} and \eqref{eq33}, we identify the effective dimension $d_{\text{eff}}$ of the reachable state manifold: $d_{\text{eff}} = n_w N_{\mathrm{rg}} 2^{2k}$ for our model and $d_{\text{eff}} = N_{\mathrm{rg}} 2^{2k}$ for the pure QNN. Assuming these normalized states are approximately uniformly distributed on the unit sphere of their respective effective manifold, according to \cite{VersBook} the 
statistical overlap between any two normalized states $\ket{\Psi}$ and $\ket{\Psi'}$ with respect to an observable $O$ follows a fundamental concentration property:
\\

\textbf{Proposition 1.} 
\emph{
	Let $\ket{\Psi}$ and $\ket{\Psi'}$ be two normalized states sampled uniformly from the unit sphere of a Hilbert space with effective dimension $d_{\mathrm{eff}}$. Then, the expected value of their squared inner product is:}
\begin{align}
	\label{eq34}
	\mathbb{E}\!\left[\,|\langle \Psi | \Psi' \rangle|^2\,\right]
	= \frac{1}{d_{\mathrm{eff}}}.
\end{align}

This result bridges our covering number analysis with the \emph{Barren Plateau} phenomenon. In our study, considering a normalized state $\ket{\Psi'} = c\, O \ket{\Psi}$ with $c$ a normalization constant, and according to \eqref{eq10} and subsection~\ref{subsec21}, the gradient of the residual cost function can be obtained by PSR measurements of Pauli operators. In a similar way, if $\|O\|$ is bounded, we may regard $\mathbb{E}\!\left[|\langle O\rangle|^2\right]$ as an inner product proportional to the one given in \eqref{eq34}. In this way, we can ensure that, if we use a pure QNN, the variance of the cost-function gradient vanishes exponentially with the number of qubits as follows:   
\begin{align}
	\label{eq35}
	&\mathbb{E}\left[\left(\frac{\partial \mathcal{C}}{\partial \theta_j}\right)^2\right]\in \mathcal{O}\left(\frac{1}{2^{2k}N_\mathrm{rg}}\right)
\end{align}

Note that \eqref{eq35} exhibits the typical behaviour 
$O\!\left(\frac{1}{b^{k}}\right)$ of the variance of the gradient of a measurement with respect to the number of qubits $k$, corresponding to the barren plateau problem \cite{McClean2018}--\cite{Marrero2021}. On the other hand, we can see that, if we use the proposed composition of QNN and CNN, the variance of the cost-function gradient also suffers exponential decay: 
\begin{align}
	\label{eq36}
	&\mathbb{E}\left[\left(\frac{\partial \mathcal{C}}{\partial w_\beta}\right)^2\right]\in \mathcal{O}\left(\frac{1}{n_w N_\mathrm{rg} 2^{2k}}\right)
\end{align}

Nevertheless, although our model exhibits a similar behaviour, it allows us to modulate $n_w N_{\mathrm{rg}} 2^{2k}$ through the number of the classical parameters $n_w$. This provides a mechanism to prevent the vanishing of gradients by fine-tuning $n_w$, even when the number of qubits is kept low.

\section{Results}\label{sec4}

\hspace{2em}In this section we present the numerical experiments that illustrate the behaviour of the proposed hybrid QNN--CNN architecture. First, we evaluate its ability to approximate solutions of Lindblad master equations using the residual-based cost function introduced earlier. Then, we analyse the expressivity and trainability of the model through the study of gradient variance under different cost functions, connecting these observations with the theoretical predictions derived from the covering number and effective dimension analysis. The results are organized into two subsections corresponding to these complementary aspects.

\subsection{Example for Solving Lindbland Master Equation using QNN composed by CNN}

\begin{figure}[H]
	\centering
	\includegraphics [width=14 cm]{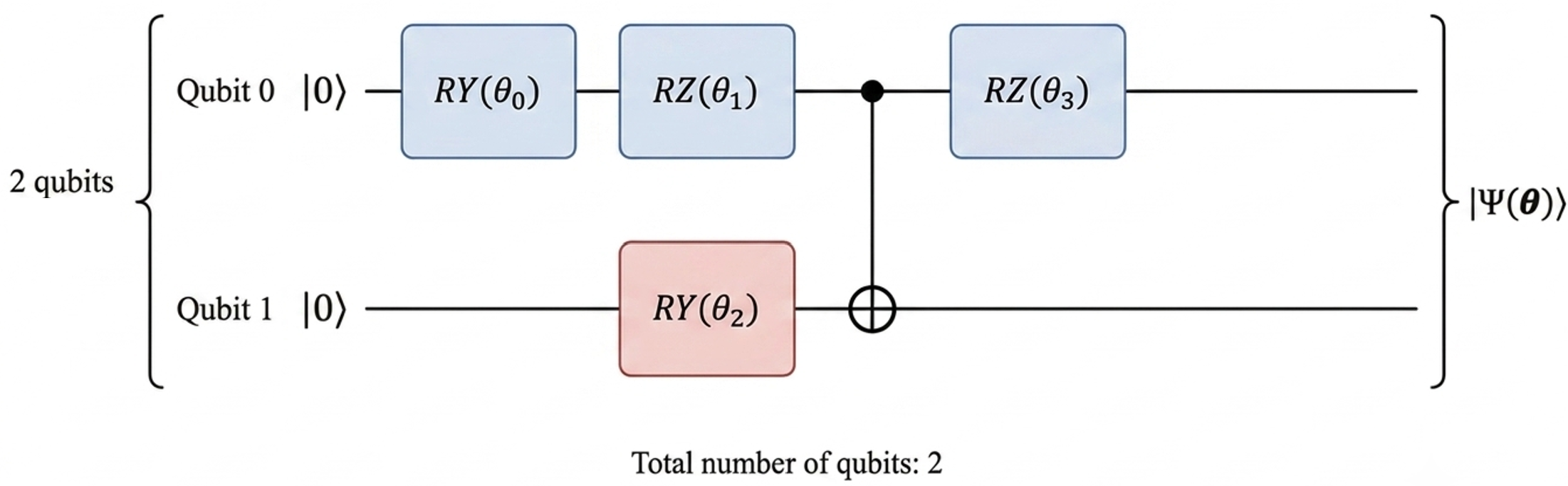}
	\caption{\label{fig3} Quantum neural network scheme of the two-qubit architecture used in the simulation.}
\end{figure}

\hspace{2em} Let us consider:
\begin{align}
	\label{eq37}
	&\mathcal{L}(\rho) = -i[H, \rho] + \gamma \left(\sigma_z \rho \sigma_z - \rho\right)
\end{align}

where $H = \frac{\omega}{2}\sigma_z$, and $O_1=\sigma_x$, $O_2=\sigma_y$, $O_3=\sigma_z$, with $\{\sigma_x,\sigma_y,\sigma_z\}$ the Pauli operators, 
$\omega = 2\pi$, and $\gamma = 1/2$. If we substitute this expression into \eqref{eq8}, we obtain:
\begin{align}
	\label{eq38}
	&\mathcal{C}(t; \mathbf{w}) = \left( \frac{d\langle\sigma_x\rangle}{dt} - (-\omega \langle\sigma_y\rangle - 2\gamma \langle\sigma_x\rangle) \right)^2+\\
	\nonumber
	&\left( \frac{d\langle\sigma_y\rangle}{dt} - (\omega \langle\sigma_x\rangle - 2\gamma \langle\sigma_y\rangle) \right)^2 + \left( \frac{d\langle\sigma_z\rangle}{dt}\right)^2
\end{align}

where $\langle \sigma_i \rangle = \mathrm{Tr}(\rho\,\sigma_i)$, with $\rho$ given by \eqref{eq5} and \eqref{eq3}, where:
\begin{align}
	\label{eq39}
	&U(\bm{\theta})=\left(RZ(\theta_3)\otimes I\right)\text{CNOT}\left(RZ(\theta_1)RY(\theta_0)\otimes RY(\theta_2)\right)
\end{align}

being $RZ(\theta_j) = e^{-\mathrm{i}\theta_j \sigma_z/2}$ and $RY(\theta_j) = e^{-\mathrm{i}\theta_j \sigma_y/2}$, with $I$ the identity operator. Figure~\ref{fig3} shows a schematic representation of the QNN used.

The rotation angles are obtained by means of \eqref{eq6}, where we fix $\bm{\theta}_0 = (0.4,\, 0.8,\, 0.2,\, 0.5)$. We minimize the cost function given in \eqref{eq38} by means of the gradient descent method \eqref{eq9}, with $\eta = 0.05$.

The differential equations obtained from the Lindblad master equation are given by:
\begin{align}
	\label{eq40}
	&\frac{d\langle\sigma_x\rangle}{dt} = -\omega \langle\sigma_y\rangle - 2\gamma \langle\sigma_x\rangle\\
	&\frac{d\langle\sigma_y\rangle}{dt} = \omega \langle\sigma_x\rangle - 2\gamma \langle\sigma_y\rangle\\  
	&\frac{d\langle\sigma_z\rangle}{dt} = 0
\end{align}

and their analytic solutions are given by:
\begin{align}
	\label{eq41}
	&\langle\sigma_x\rangle = e^{-2\gamma t}(\sigma_{x,0}\cos(\omega t) - \sigma_{y,0} \sin(\omega t))\\
	&\langle\sigma_y\rangle = e^{-2\gamma t}  (\sigma_{x,0} \sin(\omega t) + \sigma_{y,0} \cos(\omega t))\\  
	&\langle\sigma_z\rangle = \sigma_{z,0} 
\end{align}

where $\sigma_{x,0}$, $\sigma_{y,0}$ and $\sigma_{z,0}$ are obtained as the 
expected values of the Pauli operators using $\rho(\bm{\theta}_0)$ as the initial 
density matrix. Note that the resulting expressions for $\langle\sigma_x\rangle$ and $\langle\sigma_y\rangle$ are more interesting than 
$\langle\sigma_z\rangle$, since they exhibit a damped oscillator behaviour, whereas $\langle\sigma_z\rangle$ remains constant. For this reason, we focus our 
attention on capturing the behaviour of $\langle\sigma_x\rangle$ and $\langle\sigma_y\rangle$.

\begin{figure}[H]
	\centering
	\includegraphics [width=12 cm]{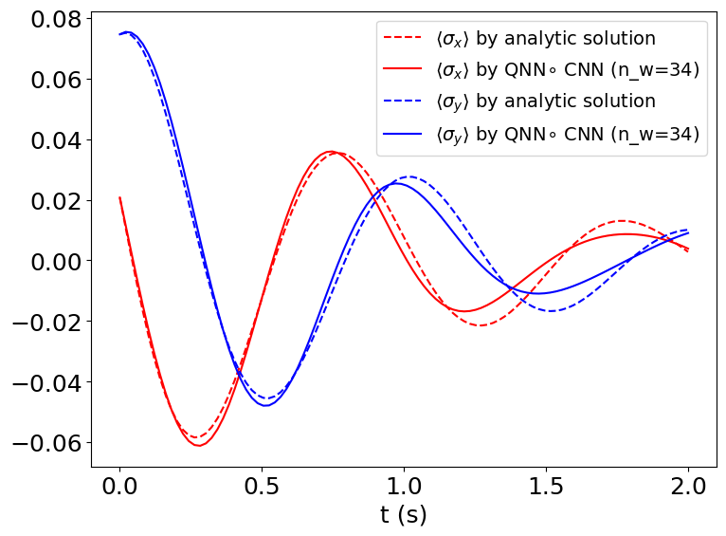}
	\caption{\label{fig4} Comparison between the expected values $\langle\sigma_x\rangle$ and 
	$\langle\sigma_y\rangle$ as functions of time $t$, obtained from the analytic solution and from the proposed model in Figure~\ref{fig3} at the final iteration for $n_w = 34$. Numerical results are those produced by the proposed model.} 
\end{figure}

Figure~\ref{fig4} shows a comparison between the expected values $\langle\sigma_x\rangle$ and $\langle\sigma_y\rangle$ as functions of time $t$, 
obtained from the analytic solution and from the proposed model in Figure~\ref{fig3}, at the final iteration of the gradient descent method for 
$n_w = 34$. We observe an acceptable agreement between the analytical and numerical results using only $n_w = 34$ (i.e., $17$ weights and biases for $N_{\mathrm{rg}} = 4$ outputs). We also point out that 800 iterations of the gradient descent method were performed.

\begin{figure}[H]
	\centering
	\includegraphics [width=12 cm]{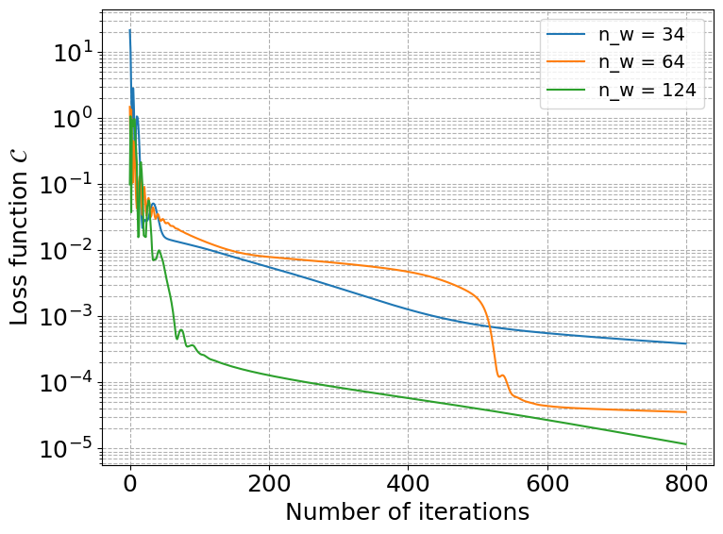}
	\caption{\label{fig5} Comparison between the values of the cost function given in \eqref{eq38} as a function of the number of iterations of the gradient descent method, obtained using $n_w = 34$, $n_w = 64$, and $n_w = 124$.} 
\end{figure}

Figure~\ref{fig5} shows a comparison between the values of the cost function given in \eqref{eq38}, as a function of the number of iterations of the gradient 
descent method, obtained using $n_w = 34$, $n_w = 64$, and $n_w = 124$. The learning rate was fixed to $\eta = 0.05$. We observe that a greater reduction in 
the cost function values is achieved as $n_w$ increases, particularly from approximately 500 iterations onward. This result indicates that it is not necessary to increase the number of qubits to capture the damped oscillator behaviour of $\langle\sigma_x\rangle$ and $\langle\sigma_y\rangle$; it is sufficient to increase the number of classical parameters $n_w$.

\begin{figure}[H]
	\centering
	\includegraphics [width=12 cm]{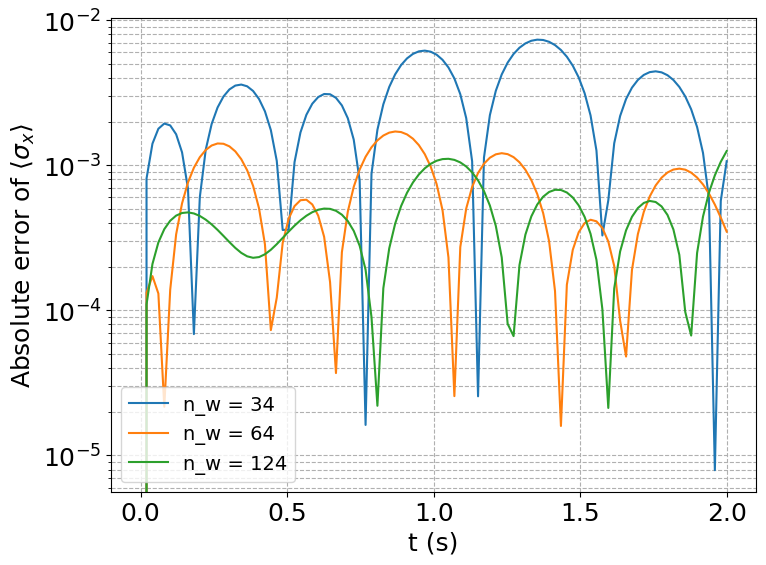}
	\caption{\label{fig6} Comparison of the absolute error between the analytic and numerical 
	expected values of $\langle\sigma_x\rangle$ and $\langle\sigma_y\rangle$ as functions of time $t$ at the final iteration, for different values of $n_w$.} 
\end{figure}

Figure~\ref{fig6} shows a comparison of the absolute error between the analytic and numerical expected values of $\langle\sigma_x\rangle$ and $\langle\sigma_y\rangle$ as functions of time $t$, for different values of $n_w$. We observe that the absolute error decreases significantly when increasing 
$n_w$ from $34$ to $64$, and decreases slightly when increasing $n_w$ from $64$ to $124$. This result confirms that increasing $n_w$ is sufficient to capture the damped oscillator behaviour of $\langle\sigma_x\rangle$ and $\langle\sigma_y\rangle$, 
without requiring an increase in the number of qubits. Therefore, the training procedure described in the previous section is suitable for this task. 

\subsection{Example on Expressivity and the Barren Plateau Problem}

\hspace{2em}In this subsection, we compute the variance of the gradient of the expected values of a global observable $O = \ket{0\ldots 0}\bra{0\ldots 0}$ and of the Lindblad residual cost function defined in\eqref{eq8}. Specifically, the residual Lindblad cost function used is given by:
\begin{align}
	\label{eq42}
	&\mathcal{C}
	=
	\left[
	\frac{d}{dt}\, \langle\sigma_z^{(0)}\rangle
	-
	\left( 2\,\langle\sigma_y^{(0)}\rangle - 2\,\gamma\, \langle\sigma_z^{(0)}\rangle \right)
	\right]^{2}
\end{align}
where $\sigma_z^{(0)}$ and $\sigma_y^{(0)}$ act on the first qubit.

\begin{figure}[H]
	\centering
	\includegraphics [width=14 cm]{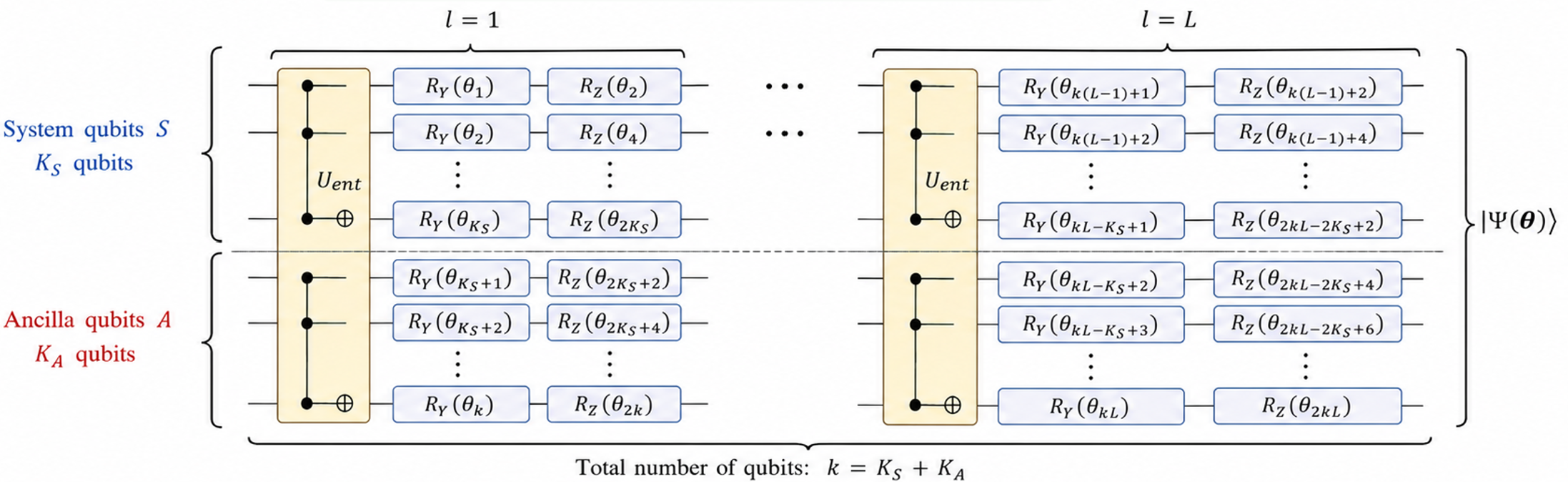}
	\caption{\label{fig7} Quantum neural network scheme used in the simulation.}
\end{figure}

First, we consider the case of a pure QNN (i.e., without composition with a classical neural network). We use the QNN defined in Section~\ref{sec2} with 
$n_\mathrm{rgq} = 2$ and CNOT gates acting between adjacent wires (see Figure~\ref{fig7}). The number of layers $L$ is fixed to the total number of qubits in order to ensure a sufficiently deep network.

\begin{figure}[H]
	\centering
	\includegraphics [width=12 cm]{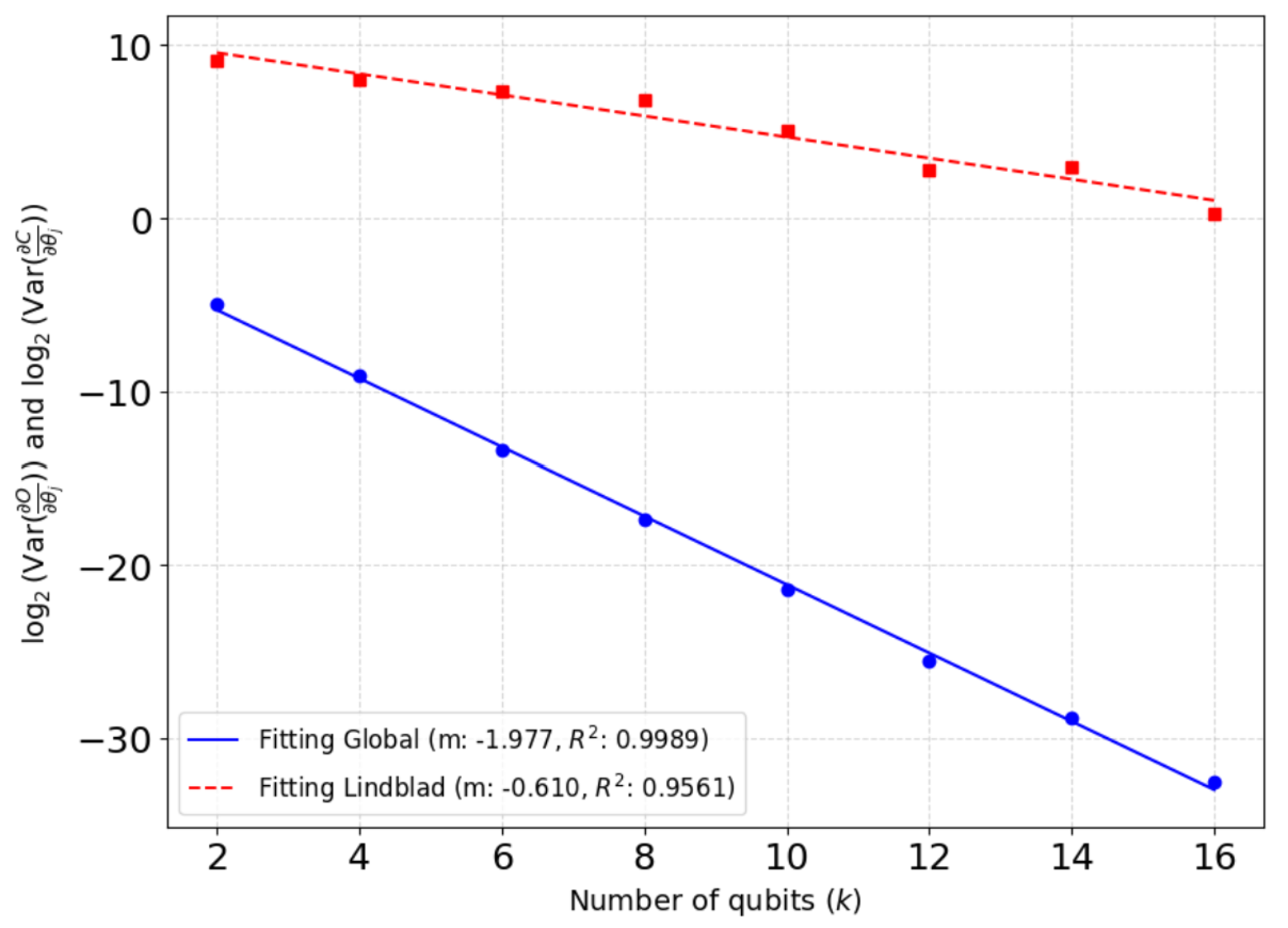}
	\caption{\label{fig8} Logarithm of the variance of the gradient with respect to the rotation angles for the expected values of a global observable and for the Lindblad residual cost function defined in \eqref{eq42}, as a function of the number of qubits. Linear least-squares fits are shown for the global observable (solid line) and for the Lindblad residual cost function (dashed line).}
\end{figure}

Figure~\ref{fig8} shows the logarithm of the variance of the gradient with respect to the rotation angles of the pure QNN shown in Figure~\ref{fig7}, as a function of the number of qubits. Results are presented for the expected values of a global observable and for the Lindblad residual cost function given in \eqref{eq42}. The variance is computed from 50 samples, each corresponding to a different set of rotation angles drawn from a uniform distribution.

We observe that an exponential decay of the variance with respect to the number of qubits appears when the global observable is used. This behaviour is not 
surprising, since the use of a deep network naturally leads to the barren plateau phenomenon \cite{McClean2018}. Moreover, a linear fit obtained via the least squares method shows a linear decay of the logarithm of the variance with slope $-1.977$ and coefficient of determination $r^2 = 0.9989$. According to \eqref{eq35}, a slope close to $-2$ is expected. 

In the case of the Lindblad residual cost function, the logarithm of the variance of the gradient exhibits a linear decay with a significantly small slope of 
$0.610$. In this way, the barren plateau problem is mitigated when using the Lindblad residual cost function. According to \cite{Cerezo2021}, barren plateaus can be alleviated when local observables acting only on a subset of qubits are employed. Therefore, this behaviour is not surprising, since the Lindblad cost function contains local observables such as $\sigma_z$ and $\sigma_y$.
\\

\begin{figure}[H]
	\centering
	\includegraphics [width=13 cm]{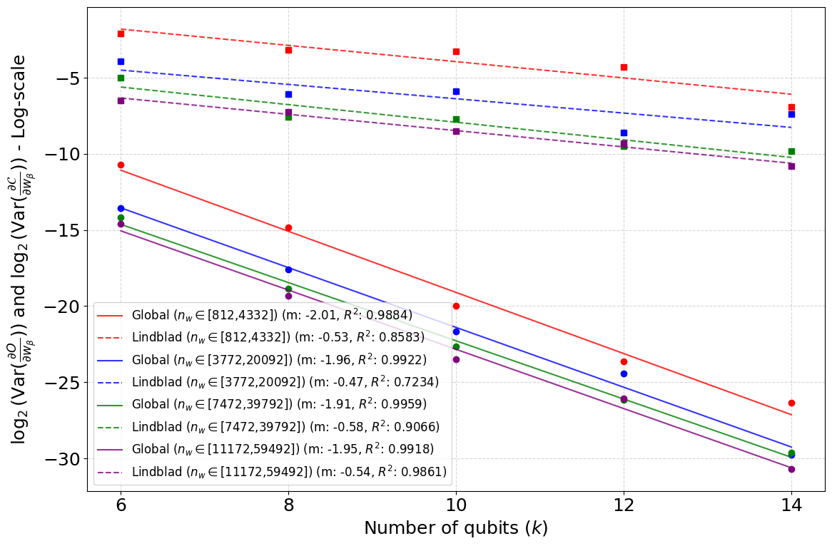}
	\caption{\label{fig9} Variance of the gradient with respect to the CNN parameters $\mathbf{w}$ as a function of the number of qubits, for the expected values of a global observable and for the Lindblad residual cost function defined in \eqref{eq42}. Results are shown for different intervals of $n_w$. Linear least-squares fits are shown for the global observable (solid lines) and for the Lindblad residual cost function (dashed lines).}
\end{figure}

Now, let us consider the same QNN shown in Figure~\ref{fig7}, but with rotation angles given by \eqref{eq6} and \eqref{eq7} (i.e., a QNN composed with a CNN). Figure~\ref{fig9} shows the logarithm of the variance of the gradient with respect to the CNN parameters $\mathbf{w}$, as a function of the number of qubits. In this case, the variance is computed from 50 samples, each corresponding to a different set of classical parameters $\mathbf{w}$ drawn from a uniform 
distribution. As before, results are shown for the expected values of a global observable and for the Lindblad residual cost function defined in \eqref{eq42}, but now evaluated at different values of $n_w$.

Note that different intervals of $n_w$ have been considered. This choice is motivated by the circuit structure in Figure~\ref{fig7}: increasing the number 
of qubits requires a larger number of rotation angles, and consequently a larger output dimension of the classical network. Therefore, to ensure a suitable 
modelling capacity, increasing the number of qubits naturally leads to an increase in $n_w$.

Again, a linear fit obtained via the least squares method shows a linear decay of the logarithm of the variance with respect to the number of qubits, with slopes approximately equal to $-2$ for the global observable and approximately $-0.5$ for the Lindblad residual cost function defined in \eqref{eq42}. These results are consistent with the expected behaviour given in \eqref{eq36} for the global observable and with the mitigation mechanism described in \cite{Cerezo2021} for the Lindblad residual cost function. In other words, the barren plateau 
phenomenon appears when using a global observable, and is mitigated when using a local residual cost function.

\begin{figure}[H]
	\centering
	\includegraphics [width=13 cm]{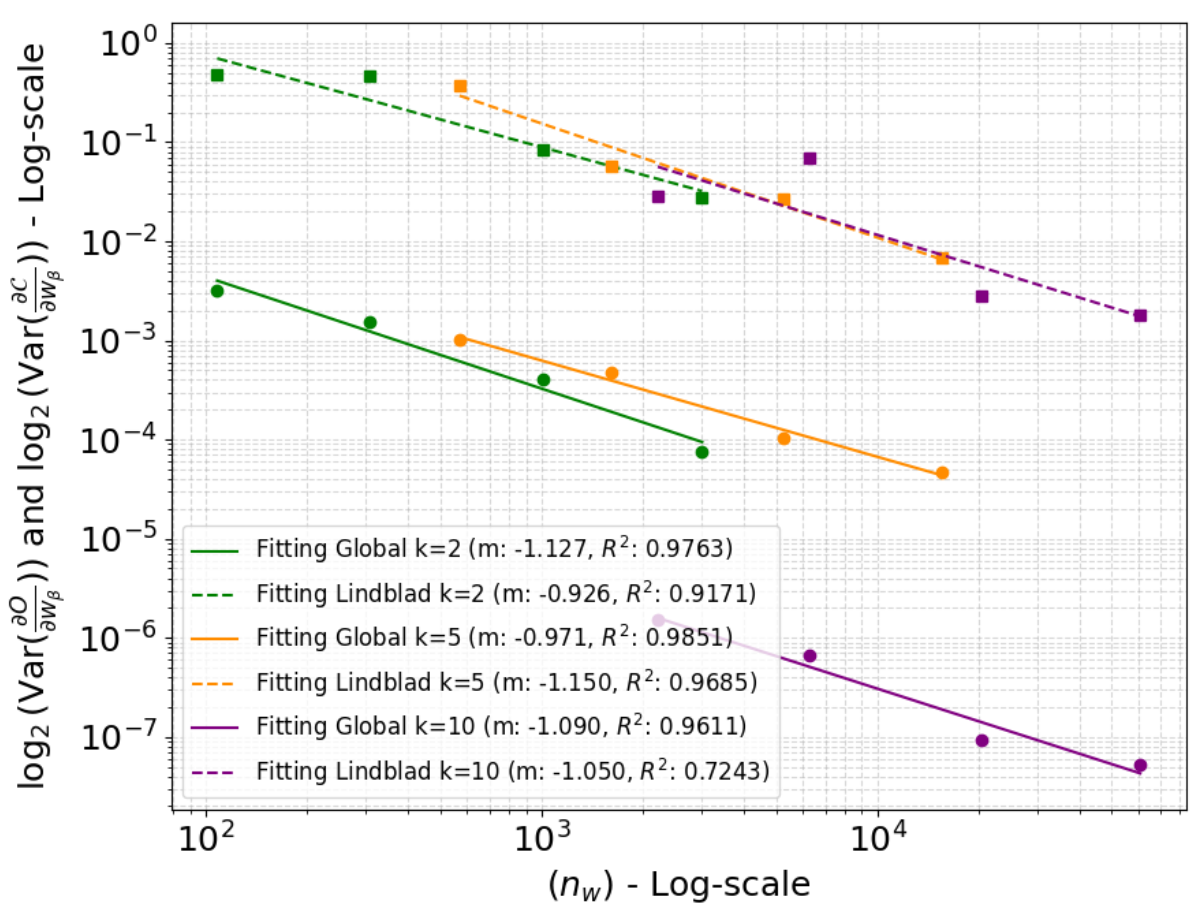}
	\caption{\label{fig10} Variance of the gradient with respect to the CNN parameters $\mathbf{w}$, in logarithmic scale, as a function of $n_w$ for different numbers of qubits. Results are shown for both cases: the global observable and the Lindblad residual cost function. Linear least-squares fits are shown for the global observable (solid lines) and for the Lindblad residual cost function (dashed lines).}
\end{figure}

Figure~\ref{fig10} shows the variance of the gradient with respect to the CNN parameters $\mathbf{w}$, in logarithmic scale, as a function of $n_w$, also in 
logarithmic scale, for different numbers of qubits. As before, results are presented for both cases: the global observable and the Lindblad residual cost 
function. A linear fit via the least squares method has been carried out for each value of the number of qubits considered. We observe that the slopes of the 
linear decay are approximately $-1$, with acceptable coefficients of determination. This behaviour is consistent with the expected decay with respect 
to $n_w$ predicted by \eqref{eq36}.
Moreover, when increasing the number of qubits from $k = 5$ to $k = 10$, we observe a reduction of at least two orders of magnitude in the variance when the 
global observable is used. This reduction does not occur when the Lindblad residual cost function is employed. In other words, the barren plateau phenomenon appears when using a global observable, whereas it is mitigated when using a local residual cost function.

\section{Conclusions}\label{sec5}

\hspace{2em} In this work, we have developed a variational framework for minimizing Lindblad's residual using an ansatz based on quantum neural networks (QNNs) whose rotation parameters are generated by a classical neural network (CNN). A central contribution of this study is the demonstration that the gradients of the Lindblad residual cost function can be computed exactly and efficiently via the parameter-shift rule, enabling the use of quantum hardware for gradient evaluation.

Beyond the optimization aspects, we have analysed the expressivity of the proposed hybrid architecture through its covering number. By comparing the  upper bound obtained for our model with the covering number of pure QNNs, we have shown that the effective dimension of the reachable state manifold expands by a factor proportional to the number of classical parameters $n_w$. This reveals that the classical network acts as a controllable mechanism that  smoothly deforms the quantum parameter space, allowing the model to represent a richer family of functions without requiring an exponential increase in the number of qubits. In particular, the degrees of freedom of the model grow linearly with $n_w$, whereas in pure QNNs they grow exponentially with the 
number of qubits. This leads to a practical training strategy: first increase $n_w$ until the desired behaviour is captured, and only then increase the number of qubits if necessary.

The covering number analysis also provides a theoretical bridge to the barren plateau phenomenon. For pure QNNs, the variance of the gradient decays exponentially with the number of qubits, as predicted by the concentration properties of high-dimensional manifolds. In contrast, the hybrid QNN--CNN architecture allows modulating the effective dimension through $n_w$, thereby preventing the complete vanishing of gradients even when the number of qubits is kept low. This theoretical insight is fully consistent with the numerical experiments: global observables lead to exponential decay of the gradient variance, while the Lindblad residual cost function---built from local observables---mitigates barren plateaus both in pure and hybrid settings.

Overall, the study shows that hybrid QNN--CNN architectures provide a flexible, expressive, and scalable approach for minimizing Lindblad's residual and approximating dissipative quantum dynamics. The ability to compute exact gradients via the parameter-shift rule, together with the controllability offered by the classical parameters, results in favourable optimization landscapes and improved expressivity. These findings support the use of hybrid variational models as a promising tool for quantum simulation on noisy intermediate-scale quantum devices.

\section*{Declarations}

\begin{itemize}
\item Funding: This work was supported by the 'Plan Estatal de Investigaci\'on', under the project PID2022-138144NB-I00 entitled 'Simetr\'ia y coherencia en sistemas cr\'iticos cu\'anticos'.
\item Conflict of interest/Competing interests: Not applicable
\item Ethics approval and consent to participate: Not applicable
\item Consent for publication: Not applicable
\item Data availability: Not applicable 
\item Materials availability: Not applicable
\item Code availability: Available under a reasonable request to the authors. 
\item Author contribution: Not applicable
\end{itemize}


\end{document}